\documentclass[galaxies,article,submit,moreauthors]{mdpi}

\firstpage{1} 
\pubvolume{xx}
\issuenum{1}
\articlenumber{5}
\pubyear{2019}
\copyrightyear{2019}
\externaleditor{Academic Editors: A.~S.~Mao and A.~Fletcher}
\history{Received: date; Accepted: date; Published: date}

\pdfoutput=1
\preto{\abstractkeywords}{\nolinenumbers}

\usepackage{color}
\usepackage[normalem]{ulem}
\usepackage{cancel} 
\definecolor{refkey}{rgb}{1,1,1}
\definecolor{labelkey}{rgb}{1,0,0}	

\usepackage{amssymb}	
\newcommand{\p}{\,{\rm pc}}
\newcommand{\kpc}{\,{\rm kpc}}
\newcommand{\km}{\,{\rm km}}
\newcommand{\kms}{\,{\rm km\,s^{-1}}}

\newcommand{\yr}{\,{\rm yr}}
\newcommand{\s}{\,{\rm s}}
\newcommand{\mkG}{\,\mu{\rm G}}
\newcommand{\cm}{\,{\rm cm}}
\newcommand{\const}{\text{const}}
\newcommand{\ion}[2]{\text{#1}{\textsc{#2}}}
\newcommand{\dd}{\mathrm{d}}
\def\vec#1{\ensuremath{\mathchoice{\mbox{\boldmath$\displaystyle#1$}}
		{\mbox{\boldmath$\textstyle#1$}}
		{\mbox{\boldmath$\scriptstyle#1$}}
		{\mbox{\boldmath$\scriptscriptstyle#1$}}}}

\newcommand{\bra}[1]{\langle #1\rangle}

\newcommand{\vv}{{\vec{v}}}

\newcommand\deriv[2]{\displaystyle\frac{\partial #1}{\partial #2}}
\newcommand{\EQ}{\begin{equation}}
\newcommand{\EN}{\end{equation}}
\newcommand{\EQA}{\begin{eqnarray}}
\newcommand{\ENA}{\end{eqnarray}}
\newcommand{\nab}{{{\nabla}}}

\newcommand{\crit}{_\text{c}} 
\newcommand{\dif}{_\text{d}} 
\newcommand{\Dloc}{\mathcal{D}} 

\Title{The Origin of Large-Scale Magnetic Fields in Low-Mass Galaxies}

\Author{Prasanta Bera $^{1,2,3\dagger}$*\orcidA,\ Anvar Shukurov 
$^{4,\dagger,}$*\orcidB\ and Kandaswamy Subramanian$^{1,\dagger}$\orcidC}

\AuthorNames{Prasanta Bera, Anvar Shukurov and Kandaswamy Subramanian}

\address{%
$^{1}$ \quad Inter-University Centre for Astronomy and Astrophysics, Post bag 
4, Pune, 411 007, India; pbera.phy@gmail.com, kandu@iucaa.in\\
$^{2}$ \quad National Centre for Radio Astrophysics, TIFR, Post Bag 3, Pune 411 
007, India;\\
$^{3}$ \quad Mathematical Sciences and STAG Research Centre, University of Southampton, Southampton SO17 1BJ, U.K.;\\
$^{4}$ \quad School of Mathematics, Statistics and Physics, Newcastle 
University, Newcastle upon Tyne, NE1 7RU, U.K.; anvar.shukurov@ncl.ac.uk}

\corres{Correspondence: pbera.phy@gmail.com, anvar.shukurov@ncl.ac.uk}

\firstnote{These authors contributed equally to this work.}

\abstract{
The origin of large-scale magnetic fields, detected in some 
low-mass (dwarf and irregular) galaxies via polarised synchrotron 
emission and Faraday rotation, remained unexplained for 
a long time. We suggest that mean-field dynamo can be active in 
galaxies of this class despite their slow rotation because their 
discs are relatively thick. Earlier assessments of the possibility 
of the mean-field dynamo action in low-mass galaxies relied on
estimates applicable to \textit{thin} discs, such as those in 
massive spiral galaxies. Using both order-of-magnitude estimates 
and numerical solutions, we show that the strength of differential
rotation required to amplify magnetic field reduces as the aspect 
ratio of the galactic gas layer increases. Thus, the puzzle of the
origin of large-scale magnetic fields appears to be solved. As a
result, this class of galaxies provides a new ground for testing
our understanding of galactic magnetism.
}
\begin{document}

\section{\label{Intro}Introduction}
Large-scale magnetic fields are a prominent and widespread in disc galaxies 
\citep{Ruzmaikin+1988p,Beck+1996,Beck15}. Their origin is most likely due to 
the mean-field dynamo action that relies on random (turbulent) motions of the 
partially ionised interstellar gas, galactic differential rotation and 
stratification of the gas distribution in the galactic gravity field. The 
mean-field dynamo theory 
\citep[for a review, see Ref.][]{Brandenburg+Subramanian2005} explains many salient
properties of galactic magnetic fields \citep{Sh07} 
\citep[for a compendium of basic results, see Ref.][]{CSSS14}. Galactic dynamo theory 
has been applied, with a significant success, to a number of specific spiral
galaxies, including the Milky Way, M31, M33, M51 \citep{SS19} and a sample of 
barred galaxies \citep[Ref.][and references therein]{BFSSSEMS05}. Statistical 
comparisons of theory and observations of galactic magnetic fields can be 
found in \citet{VEBSF15} and \citet{CST16}. \citet{RSFB15} \citep{RCSBT19} discuss galactic mean-field dynamos in the context of galaxy formation theory.

Applications of the mean-field dynamo theory to galaxies are
facilitated by the fact that the discs of spiral galaxies are thin (beyond 
a few hundred parsecs from the rotation axis) as the ratio of their scale 
height to radius is much less than unity, $\epsilon=h/r\ll1$. Near the 
Sun, this ratio is about 0.06 for the warm interstellar gas 
($h\simeq0.5\kpc$) and the galactocentric distance $r=8.5\kpc$. Asymptotic expansion in $\epsilon$ reduces the mean-field dynamo equations 
to tractable forms that can be efficiently analysed using both analytical 
and numerical methods.

Dynamo action is a threshold phenomenon: for a magnetic field to be sustained 
by plasma motions, the strength of magnetic induction effects due to 
large-scale and turbulent flows has to exceed the intensity of dissipation.
In the case of a mean (large-scale) magnetic field, the main dissipation 
mechanism is the turbulent magnetic diffusion. The intensity of the 
mean-field dynamo action is measured by the dynamo number $D$ that can be 
expressed (in a simplified form) in terms of observable quantities as 
\citep[for details, see Ref.][]{Ruzmaikin+1988p}
\begin{equation}\label{D}
D=\frac{\alpha_0\Omega S h^3}{\beta^2}\simeq9\frac{h^2\Omega S}{v_0^2}\,,
\end{equation}
where $\alpha_0\simeq l_0^2\Omega/h$ is a measure of the mean helicity of the
turbulent motions, $\Omega$ is the angular velocity of galactic rotation, 
$S=r\partial\Omega/\partial r$ is the shear rate of the rotation 
($r$ is the galactocentric distance), $h$ is the scale height 
of the gas disc, $\beta\simeq\tfrac13 l_0v_0$ is the turbulent magnetic 
diffusivity, and $l_0$ and $v_0$ are the turbulent scale and root-mean-square 
speed. (We note that $D$ is negative in the main parts of galaxies since  $\partial\Omega/\partial r<0$, so $S<0$.) 
A central result of the theory of mean-field dynamo in a thin disc is 
the marginal value of the dynamo number required for the dynamo action,
$D\crit\approx-10$: the magnetic field can grow and be maintained 
by interstellar turbulence and rotation when the magnitude of the dynamo number is large enough, $|D|>|D\crit|$, and decays otherwise.
We stress -- this will be essential later -- that this value of $D\crit$ only 
applies to a \textit{thin} disc, $\epsilon=h/r\ll1$.

Despite successes of the galactic dynamo theory, there is a class of galaxies 
whose large-scale magnetic fields it was unable to explain, namely the low-mass 
(dwarf and irregular) galaxies. In some but not all nearby dwarf galaxies, 
radio observations of synchrotron polarisation and Faraday rotation have 
revealed magnetic fields ordered on scales comparable to the galaxy size and 
exceeding the turbulent scales. Their strength is similar to, albeit often
weaker than, that in normal galaxies, a few microgauss -- e.g., in NGC~4449 
\citep{Chyzy+2000}, the Large and Small Magellanic Clouds \citep{H+91,Mao+08,Mao+12}, and NGC~1569 \citep{Kepley+10}.  \citet{Chyzy+2011}  detected synchrotron 
emission indicating random magnetic fields in three (IC~10, NGC~6822, and IC~1613) out of twelve Local Group irregular and dwarf irregular galaxies, with the upper limits of $3\text{--}5\mkG$ for the total magnetic fields in the remaining cases. \citet{HKBHWHZRR18} detected radio emission at $\lambda6\cm$ from 22 out of 40 nearby dwarf galaxies to find the average strength of the total magnetic field of 5--$8\mkG$ (comprising both random and possible large-scale parts) assuming energy equipartition between cosmic rays and magnetic fields. \citet{RCSBT19} find that 20--50\% of low-mass galaxies might host detectable large-scale magnetic fields, in agreement with \citet[][see arXiv:1302.5663 for an up-to-date list of galaxies ]{BW13}.

In comparison with large spiral galaxies, dwarf and irregular galaxies have lower 
masses and correspondingly shallower gravitational potential wells. As a 
result, they have lower rotation speeds of order $50\text{--}100\kms$ or less, 
but a similarly intense turbulence \citep[see, e.g., Table~3 in Ref.][and references 
therein]{Chyzy+2011}. For example, the maximum rotation speed of IC~2574 and 
IC~10 is about $66\kms$ and $40\kms$, respectively 
\citep{WaBr99,Wilcots+Miller1998}. Rotation is often closer to solid-body than in 
spiral galaxies \citep{Oh+15,IFNDTRB17}; as an extreme example, the maximum rotation speed 
in IC~2574 is reached at the galactocentric distance of about $8\kpc$ \citep{MCR94,WaBr99}, 
suggesting a 
typical angular velocity $\Omega\lesssim10\kms\kpc^{-1}$. Typical mean circular velocities in a sample of 17 dwarf irregular galaxies \citep{IFNDTRB17} range from  16 to $63\kms$, their \ion{H}{i} velocity dispersion is 8--$20\kms$, and the radius of the observable \ion{H}{i} disc is 0.4--$6.4\kpc$. The 26 dwarf galaxies observed in \ion{H}{i} by \citet{Oh+15} have maximum rotational velocities of 12--$126\kms$ reached at galactocentric distances 0.3--$10\kpc$.  

Residing in a shallower gravitational potential, the gas discs of dwarf 
galaxies are 3--4 times thicker than in massive spirals. In IC~2574, the 
\ion{H}{i} scale height is estimated as $h_{\ion{H}{i}}\simeq350\p$ (and the \ion{H}{i} volume density of $0.15\cm^{-3}$), whereas the one-dimensional 
velocity dispersion is $7\kms$. Similarly, $h_{\ion{H}{i}}\simeq625\p$ in 
Ho~II, and $h_{\ion{H}{i}}\simeq460\p$ in NGC~5023 \citep{WaBr99}. The aspect 
ratio of the neutral hydrogen discs of dwarf galaxies is about 10:1 as opposed 
to 100:1 in massive disc galaxies \citep{WaBr99}. The mean \ion{H}{i} axial ratio 
among 36 low surface brightness and blue compact galaxies is 0.58 \citep{S-SDK92} -- these are 
not flat objects. \citet{Roychowdhury+2013} estimate the axial 
ratio to be about $\epsilon=0.5$ for the faintest dwarf galaxies. As in spiral 
galaxies, their gas discs are flared at galactocentric distances exceeding 
$(3\text{--}4)R_\star$, where $R_\star$ is the radial scale length of the
stellar disc, $R_\star=0.8,$ 1.2, 2.1 and $0.5\kpc$ in the four dwarf galaxies 
modelled by \citet{BaJoBrBa11}, DDO~154, Ho~II, IC~2574 and NGC~2366, 
respectively.

Because of the slow rotation, the typical magnitude of the dynamo number 
\eqref{D} in the low-mass galaxies is of order unity,
\begin{equation}\label{Dd}
D\simeq 2\left(\frac{\Omega S}{100\km^2\s^{-2}\kpc^{-2}} \right)
            \left(\frac{h}{0.5\kpc} \right)^2
            \left(\frac{v_0}{10\kms} \right)^{-2}\,,
\end{equation}
significantly smaller by magnitude than $|D\crit|\approx10$ required for the mean-field dynamo 
action in a thin gas layer. This has prompted suggestions that magnetic
fields observed in dwarf and irregular galaxies are due to the fluctuation 
dynamo action that does not require any overall rotation and generates random 
magnetic fields with vanishing mean 
\citep{ZRS90,Brandenburg+Subramanian2005,SS19}. To explain the polarisation of 
observed radio emission, it is further suggested that differential rotation 
makes the random field anisotropic. However, the rotational velocity shear $S$ 
is also weak in low-mass galaxies, and such an option does not appear 
appealing. Moreover, there is little doubt that at least some such galaxies 
host genuine mean magnetic fields that produce significant Faraday rotation. As 
in other galaxies, primordial magnetic field is not a viable option 
because of the high intensity of turbulent motions.

The dominant dissipation mechanism of the large-scale magnetic field in a thin 
layer is turbulent diffusion \textit{across} it, $\tau\dif\simeq h^2/\beta$, 
where $\beta\simeq\tfrac13 l_0v_0$ is the turbulent magnetic diffusivity, 
$l_0$ is the turbulent scale and $h$ is the half-thickness (scale height) of 
the warm ionised gas layer. In spiral galaxies with $h\simeq0.5\kpc$ and 
$\beta\simeq10^{26}\cm^2\s^{-1}$, we have $\tau\dif\simeq5\times10^8\yr$. This time 
scale is larger in a thick disc, which reduces the relative importance of 
dissipation as compared to the magnetic induction effects and, as we argue 
below, the mean-fled dynamo is easier to excite in a thick layer than in a thin one. 
The magnitude of the critical dynamo number $D\crit$ increases with the layer thickness but the dynamo number increases even faster, so that the ratio $D/D\crit$, a measure of the dynamo efficiency, is higher in a thicker disc (with all other parameters being the same).
Similarly to this tendency, the dynamo efficiency in spheroids increases along the sequence from a flattened spheroid to the sphere \citep{Ivers17}.

\citet{SiSoCh18} \citep[see also Refs][]{Siejkowski+2010,SO-MSBH14} present 
numerical simulations of the mean-field dynamo action in low-mass galaxies but 
adopt turbulent magnetic diffusivity of $\beta=3\times10^{25}\cm^2\s^{-1}$, 
three times lower than the more plausible value of $\beta=10^{26}\cm^2\s^{-1}$. 
The spatial resolution of the simulations,  47 or $78\p$, is close to the turbulent scale, so that turbulent 
motions are not resolved and the adopted value of $\beta$ is likely to be close to the effective one in the simulations. As shown in Eq.~\eqref{D}, the dynamo number in those 
simulations can, thus, be overestimated by a factor $\beta^2\approx10$ and the efficiency 
of the dynamo action appears to be exaggerated, which makes the significance of these 
results questionable.

In this paper we discuss solutions of the mean-field dynamo equation in  thick 
discs to demonstrate that it can explain the amplification and maintenance of 
large-scale magnetic fields in low-mass galaxies.
Having presented basic equations in Section~\ref{BE}, we discuss dynamo action in a 
thick disc in Section~\ref{DATD} where the value of $D\crit$ is estimated and the
corresponding minimum rotational speed required for the dynamo action in a thick disc is derived.
Numerical solutions of the dynamo equation are presented in Section~\ref{NS} and explored in Section~\ref{R} with emphasis on the dependence of $D\crit$ and the corresponding critical rotational speed on the disc aspect ratio; the accuracy of the estimates of Section~\ref{DATD} is confirmed. The results are discussed and summarised in Section~\ref{CD}.

\section{\label{BE}Basic equations}
The generation of a mean (large-scale) magnetic field $\vec{B}$ by a random velocity field is governed by the mean-field dynamo equation 
\citep{Moffatt78,Parker79,KR80,ZRS83}
\begin{equation}\label{mean}
\deriv{\vec{B}}{t}=\nabla\times \left[ \vec{V} \times \vec{B} + \alpha \vec{B} -\beta
\nabla\times\vec{B}\right],
\end{equation}
where $\vec{V}=\vec{\Omega}\times\vec{r}$ is the large-scale velocity 
dominated by rotation with the angular velocity $\vec{\Omega}$ and the effects of galactic random flows (turbulence) are represented by the turbulent transport coefficients $\alpha =-\tfrac13\tau \bra{\vv\cdot(\nab\times\vv)}\simeq l_0^2\Omega/h$ (where $l_0$ is the turbulent scale and $h$ is the density scale height), which depends on the mean helicity of the turbulence $\bra{\vv\cdot(\nab\times\vv)}$, and turbulent diffusivity $\beta\simeq\tfrac13\tau\bra{\vv^2}$ (assumed to be much larger than the Ohmic magnetic diffusivity), with $\vv$ the turbulent velocity field, $\tau$ its correlation time, and angular brackets denote an appropriate average over the
turbulent fluctuations.

We model a galaxy as an axisymmetric disc with a Brandt rotation curve $V(r)=r\Omega(r)$,
\begin{equation}\label{rot_curve}
 \Omega(r) = \frac{V_0}{R_0} \left[\frac{1}{3}+\frac{2}{3}\left(\frac{r}{R_0}\right)^n\right]^{-3/(2n)}\,,
\end{equation}
in terms of cylindrical polar coordinates $(r,\phi,z)$, with $n=2$, where $V_0$ and $R_0$ are the characteristic velocity and radial scale length. 
 The corresponding velocity shear rate is defined as $S=r\partial\Omega/\partial r$.
 The half-thickness (or scale height) of the ionised gas disc is denoted $h$.
For the interstellar turbulence, we assume the correlation scale 
$l_0=0.1\kpc$ and speed $v_0=10\kms$, both similar to those in spiral galaxies. 

We consider the simplest axisymmetric solutions to the mean-field equation, and so
the mean magnetic field can be written as
\begin{equation}\label{B_A}
\vec{B} = B_\phi(r, z)\hat{\vec{\phi}}+\nabla\times\left[A_\phi(r,z) \hat{\vec{\phi}}\right],
\end{equation} 
where $B_\phi$ and $A_\phi$ are the azimuthal components of the magnetic field and the vector potential, respectively, and $\hat{\vec{\phi}}$ is the unit azimuthal vector. The assumption of axial symmetry may be rather crude when applied to low-mass galaxies but it is sufficient to assess the possibility of the mean-field dynamo action, our main goal in this paper. More realistic non-axisymmetric solutions and non-axisymmetric galaxies will be discussed elsewhere.

We use dimensionless variables, here denoted with tilde, defined in terms of the characteristic values denoted with subscript zero:
\begin{equation}\label{d_less}
\begin{split}
 r&=R_0\tilde{r}\,,\qquad z= h_0\tilde{z}\,,\qquad t=(h_0^2/\beta)\tilde{t}\,,
            \qquad  B_\phi=B_0\tilde{B}_\phi\,,\\
A_\phi&=h_0 B_0\tilde{A}_\phi\,, \qquad \Omega=\Omega_0\tilde{\Omega}\,,  
            \qquad S=S_0\tilde{S}\,,\qquad 
            S_0=\Omega_0\,,\qquad \alpha=\alpha_0\tilde{\alpha}\,,
\end{split}
\end{equation}
where
\begin{equation}
\alpha_0=l_0^2\Omega_0/h_0\,, \qquad \Omega_0=V_0/R_0\,,
\end{equation}
and we drop tilde at the dimensionless variables unless specified otherwise or obvious from the context. Since we here only consider kinematic dynamo solutions that vary exponentially in time, the characteristic magnetic field $B_0$ can be chosen arbitrarily and can be used to normalise the results as convenient. 

We consider a flat gas layer with $h=h_0=\const$. Disc flaring can easily be included
\citep{Ruzmaikin+1988p} but
it does not change our conclusions regarding the possibility of dynamo action at relatively small galactocentric distances although the radial distribution of magnetic field can be affected \citep[e.g., Ref.][]{SRBHR18}. The coefficient $\alpha(r,z)$ is assumed to be factorisable, $\alpha=\alpha_0\alpha_1(z)\Omega(r)$ since $\alpha\propto\Omega(r)$, where we adopt $\alpha_1(z)=\sin(\pi z/h)$ for numerical solutions; this choice affects the results only weakly.

In terms of the dimensionless variables, the cylindrical components of the dynamo equation reduce to
\begin{align}
 \deriv{B_\phi}{t} &= -R_\omega S \deriv{A_\phi}{z} -R_\alpha\deriv{}{z}\left(\alpha\deriv{A_\phi}{z}\right) 
 + \deriv{^2 B_\phi}{z^2}
  + \epsilon^2\mathcal{L}_r(B_\phi)\,, 
  \label{eq_dynamo_norm_Bphi0}\\
 \deriv{A_\phi}{t} &= R_\alpha\alpha B_\phi + \deriv{^2 A_\phi}{z^2}
  + \epsilon^2\mathcal{L}_r(A_\phi)
  \,, \label{eq_dynamo_norm_Aphi0}
\end{align}
where
\begin{equation}\label{RaRo}
R_\alpha=\frac{h_0\alpha_0}{\beta}\,,\qquad  
R_\omega=\frac{S_0 h^2}{\beta}\,,  \qquad 
\epsilon=\frac{h_0}{R_0}\,, \qquad 
\mathcal{L}_r(F)=\deriv{}{r}\left[\frac{1}{r}\deriv{(r F)}{r}\right].
\end{equation}
and we note that $\alpha(r,z)$ and $S(r)$ are functions of position. Magnetic induction effects are quantified by dimensionless dynamo numbers $R_\omega$ for the differential rotation and $R_\alpha$ for the helical turbulence. The dynamo system is controlled by three dimensionless parameters, $\epsilon$, $V_0/v_0$ and $l_0/h_0$ via
\begin{equation}\label{RaRoep}
R_\alpha=3\epsilon\frac{l_0}{h_0}\,\frac{V_0}{v_0}\,,
\qquad R_\omega=3\epsilon\frac{h_0}{l_0}\,\frac{V_0}{v_0}\,.
\end{equation}
An advantage of selecting $R_0$ (the radial scale of the rotation curve) as the unit radial length is that the estimates and numerical results presented below are insensitive to the specific value of $R_0$ as this parameter enters the results only via the disc aspect ratio $\epsilon=h_0/R_0$.

When the term with $R_\alpha$ in Eq.~\eqref{eq_dynamo_norm_Bphi0} is neglected, the dynamo system is called the $\alpha\omega$-dynamo and its efficiency is fully characterised by the single dimensionless number, the dynamo number \begin{equation}\label{Dep}
    D=R_\alpha R_\omega=9\epsilon^2\frac{V_0^2}{v_0^2}\,.
\end{equation}
This approximation is appropriate when the differential rotation is sufficiently strong to have $R_\omega\gg R_\alpha$. Otherwise, the dynamo is known as the $\alpha^2\omega$-dynamo, characterised by two dimensionless numbers, $R_\alpha$ and $R_\omega$ or, equivalently, $R_\alpha^2$ and $D$.
For parameter values typical of low-mass galaxies, $\epsilon=0.5$, $V_0/v_0=5\text{--}10$ and $l_0/h_0=0.2$, we have
$R_\alpha=1.5\text{--}3$ and $R_\omega\simeq30\text{--}70$.

\section{\label{DATD}Mean-field dynamo in a thick disc}
As any other dynamo mechanism, the mean-field dynamo is a threshold phenomenon: the mean magnetic field can be amplified and then maintained against turbulent diffusion only if the induction effects are strong enough. When differential rotation is strong (the $\alpha\omega$-dynamo approximation), the mean magnetic field grows and then reaches a steady state if $D>D\crit$ for a certain $D\crit$ and decays if $D<D\crit$. 

In galaxies with significant differential rotation, it dominates over the $\alpha$-effect in Eq.~(\ref{eq_dynamo_norm_Bphi0}) and the term with $R_\alpha$ can be neglected in this equation. To obtain an estimate of the dynamo threshold in a thick disc of a dwarf galaxy, we note that, in terms of dimensionless variables, spatial derivatives in equations~\eqref{eq_dynamo_norm_Bphi0} and \eqref{eq_dynamo_norm_Aphi0} are of order unity in $\epsilon$. In the spirit of the no-$z$ approximation introduced by \citet{SM93} and refined by \citet{P01} \citep[details and further justification can be found in Ref.][]{JCBS14}, we have the following estimates: $|\partial B_\phi/\partial z|\simeq \pi|B_\phi|/2$, $|\partial^2 B_\phi/\partial z^2|\simeq-\pi^2|B_\phi|/4$ and $|\mathcal{L}_r(B_\phi)|\simeq -k^2|B_\phi|$, and similarly for the derivatives of $A_\phi$. The negative sign in the second derivatives follows from the properties of the diffusion operator and $k$ is a constant of order unity.

The dynamo is called marginal when $\partial\vec{B}/\partial{t}=\vec{0}$; this state corresponds to the marginal value of the dynamo number $D\crit=R_\alpha R_\omega\overline{\alpha}\overline{S}$, where $\overline{\alpha}$ and $\overline{S}$ are the values of $\alpha$ and $S$ averaged in radius (note that $\overline{S}<0$, hence $D\crit<0$, and both $\overline{\alpha}$ and $\overline{\Omega}$ are dimensionless, so of order unity). For the marginal state,  equations~\eqref{eq_dynamo_norm_Bphi0} and \eqref{eq_dynamo_norm_Aphi0} reduce, by the order of magnitude in $\epsilon$, to
\begin{equation}
\frac{\pi}{2}R_{\omega,\text{c}}\overline{S}A_\phi
+\left(\frac{\pi^2}{4}+k^2\epsilon^2\right)B_\phi\simeq0\,, \qquad
\left(\frac{\pi^2}{4}+k^2\epsilon^2\right)A_\phi
                    -R_{\alpha,\text{c}}\overline{\alpha}B_\phi\simeq0\,,
\end{equation}
leading to an order of magnitude estimate 
\begin{equation}\label{Dc2}
D\crit\simeq-\frac{2}{\pi} \left(\frac{\pi^2}{4}+k^2\epsilon^2\right)^2\,,
\end{equation}
where we adopt $\overline{\alpha}=\overline{\Omega}=1$. Combining the estimate of $D\crit$ with equation~\eqref{Dep}, we obtain a measure of the local dynamo efficiency as
\begin{equation}\label{DDcr}
\frac{D}{D\crit}\simeq \frac{9\pi}{2}\frac{V_0^2}{v_0^2}
        \frac{\epsilon^2}{(\pi^2/4+k^2\epsilon^2)^2}\,,
\end{equation}
confirming that the mean-field dynamo is more efficient in a thicker disc: this ratio increases with $\epsilon$ for $\epsilon<\pi/(2k)$, with the right-had side smaller than unity for $k>\pi/2$.

This estimate can also be rewritten in terms of the rotational velocity using 
equations~\eqref{RaRoep}: magnetic field can be maintained when the characteristic
rotation speed $V_0$ exceeds a critical value $V\crit$ given by
\begin{equation}\label{Vc}
V\crit\simeq v_0\sqrt{\frac{2}{\pi}}\, 
                            \frac{\pi^2/4+k^2\epsilon^2}{3\epsilon}\,,
\end{equation}
and $V\crit$ decreases with $\epsilon$ for $\epsilon<\pi/(2k)$. This confirms again that the mean-field dynamo is easier to excite in a thicker disc.

It is convenient to introduce the local dynamo number
\begin{equation}
\mathcal{D}=D\alpha_1(r)S(r)\,,
\end{equation}
a function of galactocentric radius that quantifies the dynamo efficiency at a given $r$. For a magnetic field growing at an exponential growth rate $\gamma$, i.e., $\partial B_\phi/\partial t = \gamma B_\phi$ and $\partial A_\phi/\partial t = \gamma A_\phi$, estimates similar to those presented above yield $\gamma\simeq \sqrt{-\pi\Dloc/2}-\sqrt{-\pi D\crit/2}$ or, in dimensional variables, the local dynamo amplification (e-folding) time is estimated, for $k\epsilon=\pi/2$, as 
\begin{equation}\label{gamma}
\gamma^{-1}\simeq\frac{3h_0^2}{l_0v_0}\,
\frac{1}{3\epsilon V_0/v_0\sqrt{-\pi S\Omega/(2\Omega_0^2)}-\pi^2/2}\simeq 10^9\yr\,,
\end{equation}
where the numerical value is obtained for $\epsilon=0.5$, $V_0/v_0=3$, $v_0=10\kms$, $h_0=0.5\kpc$, $l_0/h_0=0.2$ and $S\Omega/\Omega_0^2=-1$.

The ratio $D/D\crit$ and $\gamma$ are maximum, whereas $V\crit$ is minimum, for the mode with the radial wave number $k=2/(\pi\epsilon)$. For this leading mode, we obtain $D/D\crit\simeq (9/\pi^3)\epsilon^2(V_0/v_0)^2$ and $V\crit\simeq\pi^2 v_0/(6\epsilon)\simeq 3v_0\simeq30\kms$, with the numerical coefficient obtained for $\epsilon=0.5$.

These estimates implicitly assume the quadrupolar parity of the large-scale magnetic field since the relations like $|\partial B_\phi/\partial z|\simeq \pi|B_\phi|/2$, $|\partial^2 B_\phi/\partial z^2|\simeq-\pi^2|B_\phi|/4$ and $|\mathcal{L}_r(B_\phi)|\simeq -k^2|B_\phi|$ only apply in this case. The asymptotic solution for $|D|\ll1$ \citep{ShSo08} (not presented here) shows that dipolar modes have much higher values of $|D\crit|$ and $V\crit$ and are oscillatory.

\section{\label{NS}Numerical solutions}
We confirm these estimates using numerical solutions of the dynamo equation. Equations~\eqref{eq_dynamo_norm_Bphi0} and \eqref{eq_dynamo_norm_Aphi0} are solved within a rectangle $(0,0)\leq(r,z)\leq(R_\text{b},h_0)$ in the $(r,z)$-plane. We adopt $R_\text{b}=4$ and verified that the results are not sensitive to the position of the outer radial boundary when $R_\text{b}=3\text{--}10$. The boundary conditions adopted in radius, $\partial(rB_\phi)/\partial r = \partial(rA_\phi)/\partial r=0$ at $r=R_\text{b}$ correspond to vanishing diffusive flux of magnetic field. The boundary conditions on the disc surface, $B_\phi=\partial A_\phi/\partial z=0$ at $|z|=h$, apply when large-scale electric currents outside the disc are much weaker than those within the disc (the vacuum boundary conditions). The boundary condition for $A_\phi$ at $|z|=h$ is an approximation to $\partial A_\phi/\partial z=\mathcal{O}(\epsilon A_\phi)$, and its accuracy reduces as $\epsilon$ increases \citep[Ref.][and references therein]{Sh07}. Boundary conditions at the disc mid-plane control the symmetry (either quadrupolar/even or dipolar/odd parity) of the magnetic field across the disc \citep[see Ref.][for details]{Ruzmaikin+1988p}
\begin{equation}\label{BC}
\begin{split}
\left.\deriv{B_\phi}{z}\right|_{z=0}= A_\phi|_{z=0} = 0 &\qquad\text{(quadrupolar)}\,,\\
B_\phi|_{z=0} = \left.\deriv{A_\phi}{z}\right|_{z=0}= 0 &\qquad\text{(dipolar)}\,,
\end{split}
\end{equation}
and we are free to consider the range $z>0$ because of the symmetry of the solutions. The initial magnetic field is of a large scale, with sinusoidal variation in both $r$ and $z$ to satisfy the boundary conditions and has the average strength $|\vec{B}|\approx10^{-3}$ in dimensionless units (its magnitude does not, in fact, matter as long as kinematic dynamo solutions are concerned).

The solutions are obtained for the rotation curve \eqref{rot_curve} 
$\alpha(r,z)=\Omega(r)\sin(\pi z)$ in dimensionless units. Forward time stepping is performed using the fourth-order Runge--Kutta scheme and central finite differences are used for the spatial derivatives. The consistency of the results is 
verified by increasing the grid resolution. Here we present results for $64\times64$ and $128\times128$ grids, with a uniform spacing in both radial and vertical directions (in terms of dimensionless variables).
 
An efficient technique  \citep{RaWi89} to find the critical value of the dynamo number, $D\crit$
for a given $\alpha(r,z)$ is to solve the dynamo equations for a sufficiently large (supercritical) $R_\alpha$ with a modified form of the coefficient, $\hat{\alpha}(r,z)$, that depends on $|\vec{B}|$ (so that the field is eventually driven to a steady state) but has the same spatial form as $\alpha(r,z)$ at all times:
\begin{equation}\label{a_int}
\widehat{\alpha}(r,z,t)=\frac{\alpha_0\alpha(r,z)}{1+2\pi W^{-1}\int_0^{R_\text{b}}r\,\dd r\int_0^h \dd z\,
|\vec{B}(r,z,t)/B_0|^2}\,,
\end{equation}
where $W=\pi hR_\text{b}^2$ is the volume of the computational domain and we use 
dimensional variables. The value of $\widehat{\alpha}(r,z,t)/\alpha(r,z)$ in the steady state (when $\widehat{\alpha}$ no longer varies with $t$) for $\alpha_0$ in equations~\eqref{RaRo} and \eqref{Dep} yields $D\crit$.

\section{\label{R}Results}
Our goal is to clarify conditions required for mean-field dynamo action in the thick discs of low-mass galaxies rather than to explore magnetic field forms and distributions in specific galaxies. Therefore, we focus on finding the critical values of the dynamo number and the corresponding rotation speed and only briefly discuss steady-state magnetic field distributions.

\begin{figure}
\centering
\includegraphics[width=0.7\columnwidth]{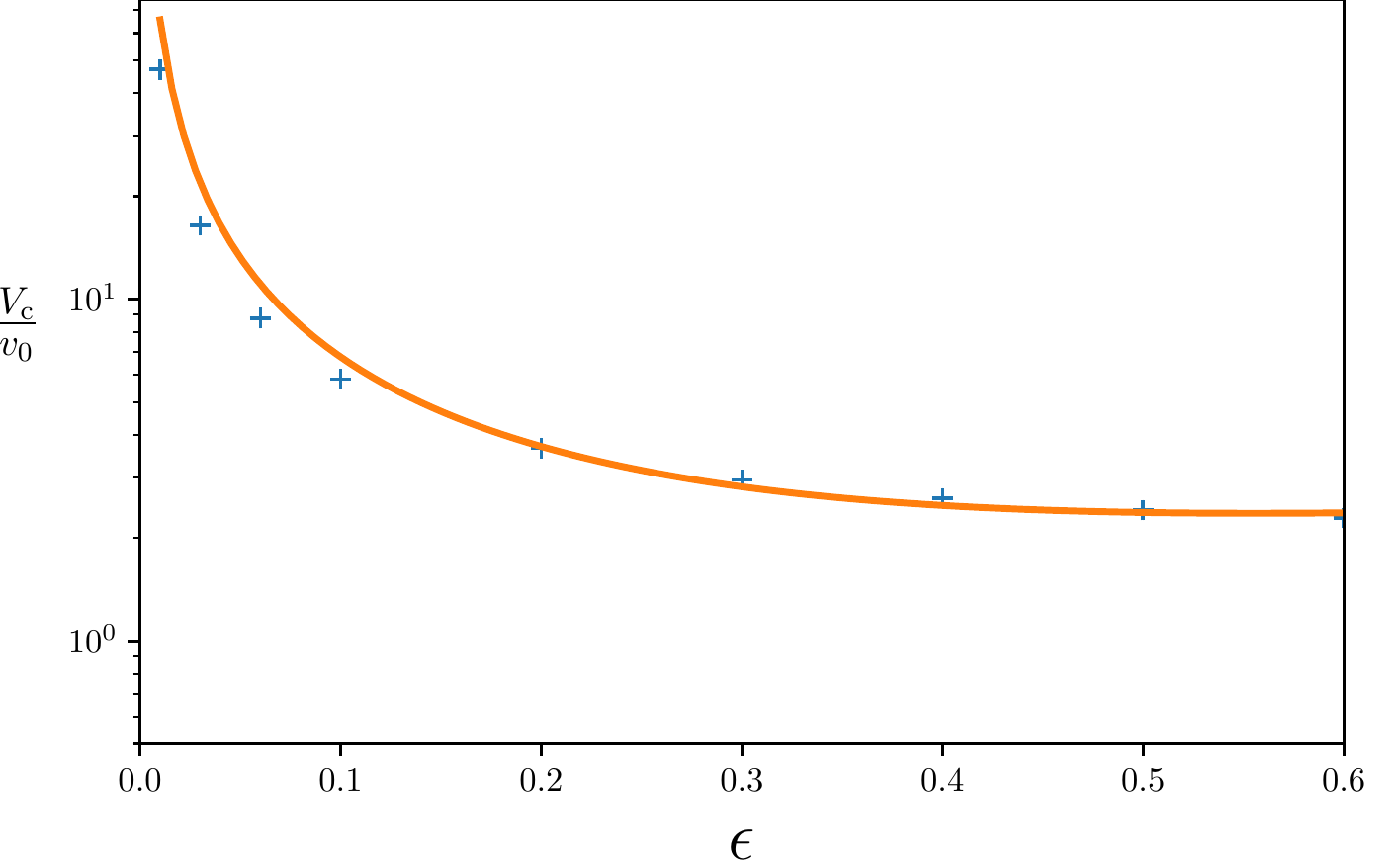}
\caption{The variation of the critical rotation speed $V\crit$ required for the dynamo action  (for the quadrupolar magnetic field geometry) with the disc aspect ratio $\epsilon$: numerical results are shown with crosses and the solid line represents the approximate value from Eq.~\eqref{Vc} with $k^2=8$.
    }
    \label{fig:Vcrit_epsilon}
\end{figure}

\begin{figure}
\centering
\includegraphics[width=0.7\columnwidth]{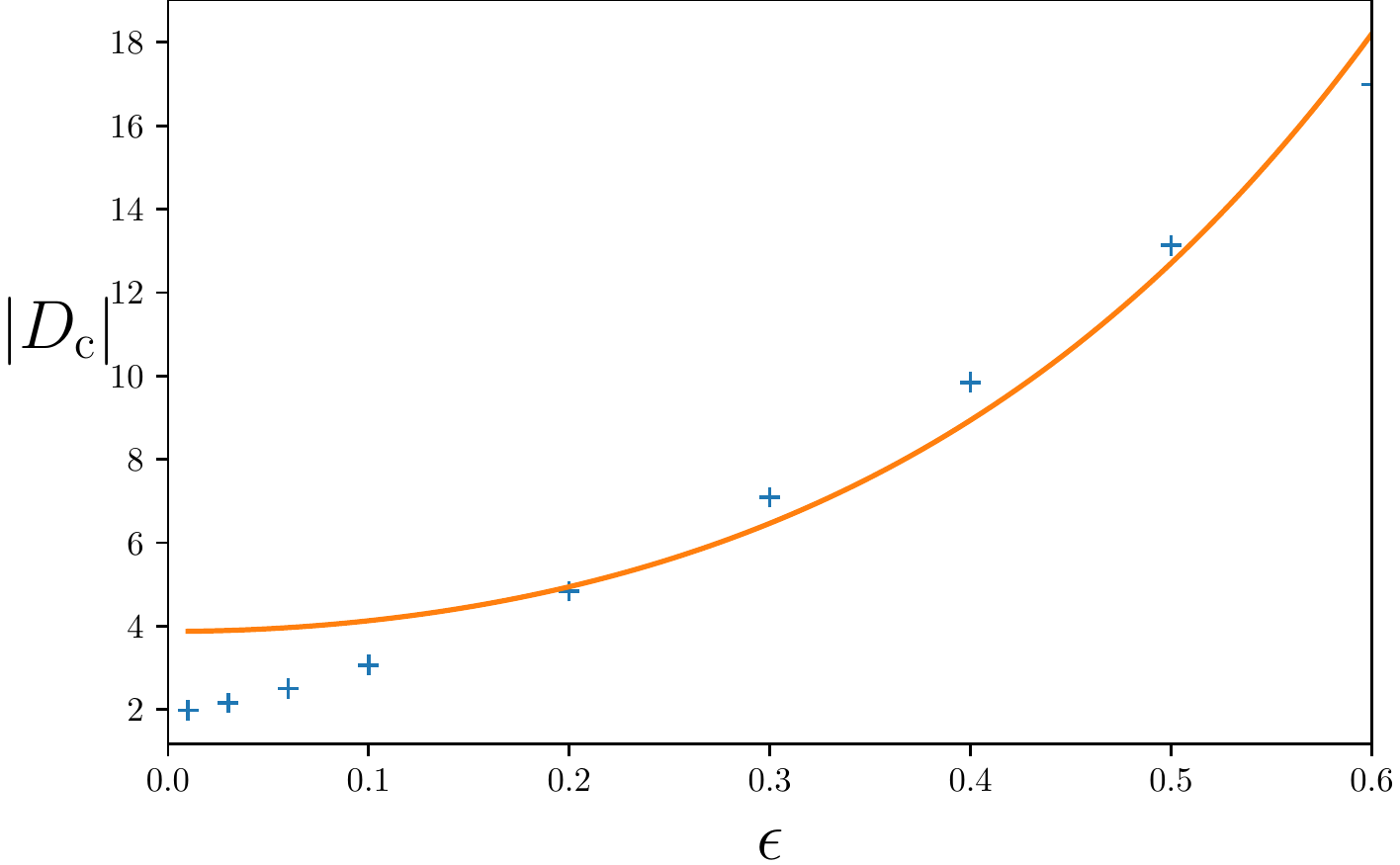}
\caption{As in Fig.~\ref{fig:Vcrit_epsilon} but for the critical dynamo number, Eq.~\eqref{DDcr}. The accuracy of the approximate solution increases noticeably with $\epsilon$.
    }
    \label{Dcr_eps}
\end{figure}

\subsection{Critical rotation speed}\label{CRS}
We solve equations~\eqref{eq_dynamo_norm_Bphi0} and \eqref{eq_dynamo_norm_Aphi0} with the vacuum boundary conditions and symmetry conditions \eqref{BC} for a range of the disc aspect ratios $\epsilon$ and a range of $R_0$ in the rotation curve \eqref{rot_curve} to obtain $D\crit$ and then the corresponding critical rotation speed is obtained which is then compared with equation~\eqref{Vc}. For  given $R_0$, the disc aspect ratio $\epsilon$ is varied by changing the characteristic disc scale height $h_0$. We recall that the large-scale magnetic field can be maintained if $D\geq D\crit$ or, equivalently, $V_0\geq V\crit$ and decays otherwise. Figure~\ref{fig:Vcrit_epsilon} shows $V\crit$ as a function of $\epsilon$; Fig.~\ref{Dcr_eps} presents the variation of $D\crit$. As expected, weaker rotation is required to excite the dynamo as the disc becomes thicker. 

It is remarkable how accurately equations~\eqref{Dc2} and \eqref{Vc} reproduce the variations of $D\crit$ and $V\crit$ with $\epsilon$. The best agreement between the approximate solution of Section~\ref{DATD} and the numerical results is achieved for $k^2\approx8$. This value of $k$ is of order unity as expected but it still exceeds unity, which indicates that the magnetic field has the radial scale of order $R_0/k$, about a third of the disc radius, in qualitative similarity to the kinematic dynamo solutions in a thin disc \citep{Ruzmaikin+1988p}.

Figure~\ref{fig:Vcrit_epsilon} confirms that even weak rotation, with the maximum rotation speed only a few times larger than the turbulent speed, is sufficient to support the mean-field dynamo action in a thick disc typical of a low-mass galaxy. For the aspect ratios close to $\epsilon=0.3\text{--}0.5$, even $V_0\simeq20\text{--}30\kms$ appears to be be sufficient to maintain a large-scale quadrupolar magnetic field. For a dipolar field in a disc of a similar aspect ratio, the critical velocity is about $100\kms$.

\begin{figure}
	\includegraphics[width=\columnwidth]{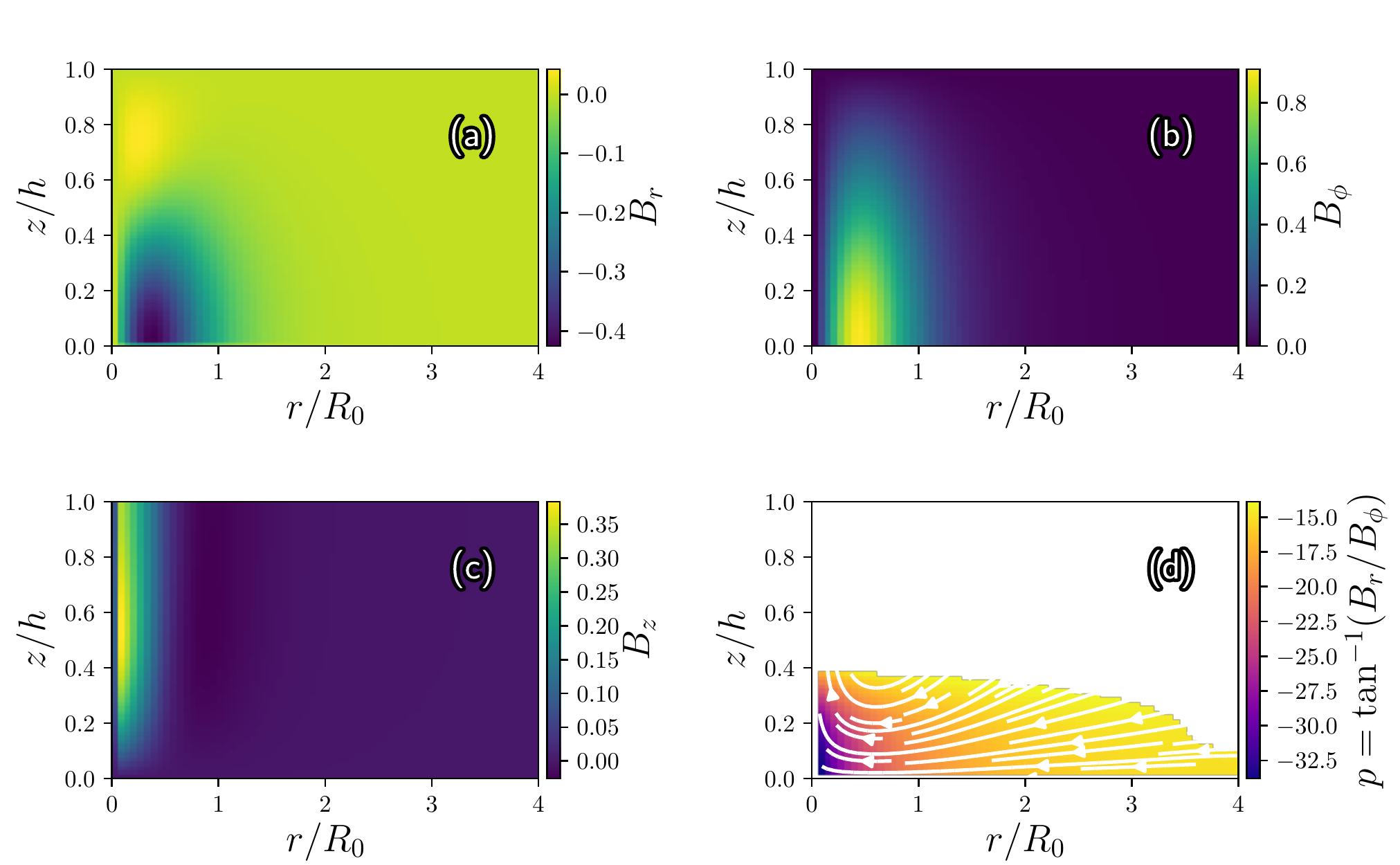}
    \caption{The cylindrical components (a) $B_r$, (b)~$B_\phi$ and (c) $B_z$ of the marginally stable quadrupolar eigenfunction of the $\alpha\omega$-dynamo equations for $R_0=1\kpc$, $V_0/v_0=2.96$ and $\epsilon=0.3$ normalised to $\max(B_r^2+B_z^2+B_\phi^2)=1$. Panel (d) shows poloidal magnetic lines (white), providing an indication of the relation between $B_z$ and $B_r$, and the pitch angle of magnetic lines (in degrees, colour).
    Only the region where $B_r^2+B_z^2+B_\phi^2\geq 0.01\max(B_r^2+B_z^2+B_\phi^2)$ is shown.\\
    }
    \label{fig:2dBcomponentsQuadrupolar}
\end{figure}

\begin{figure}
	\includegraphics[width=\columnwidth]{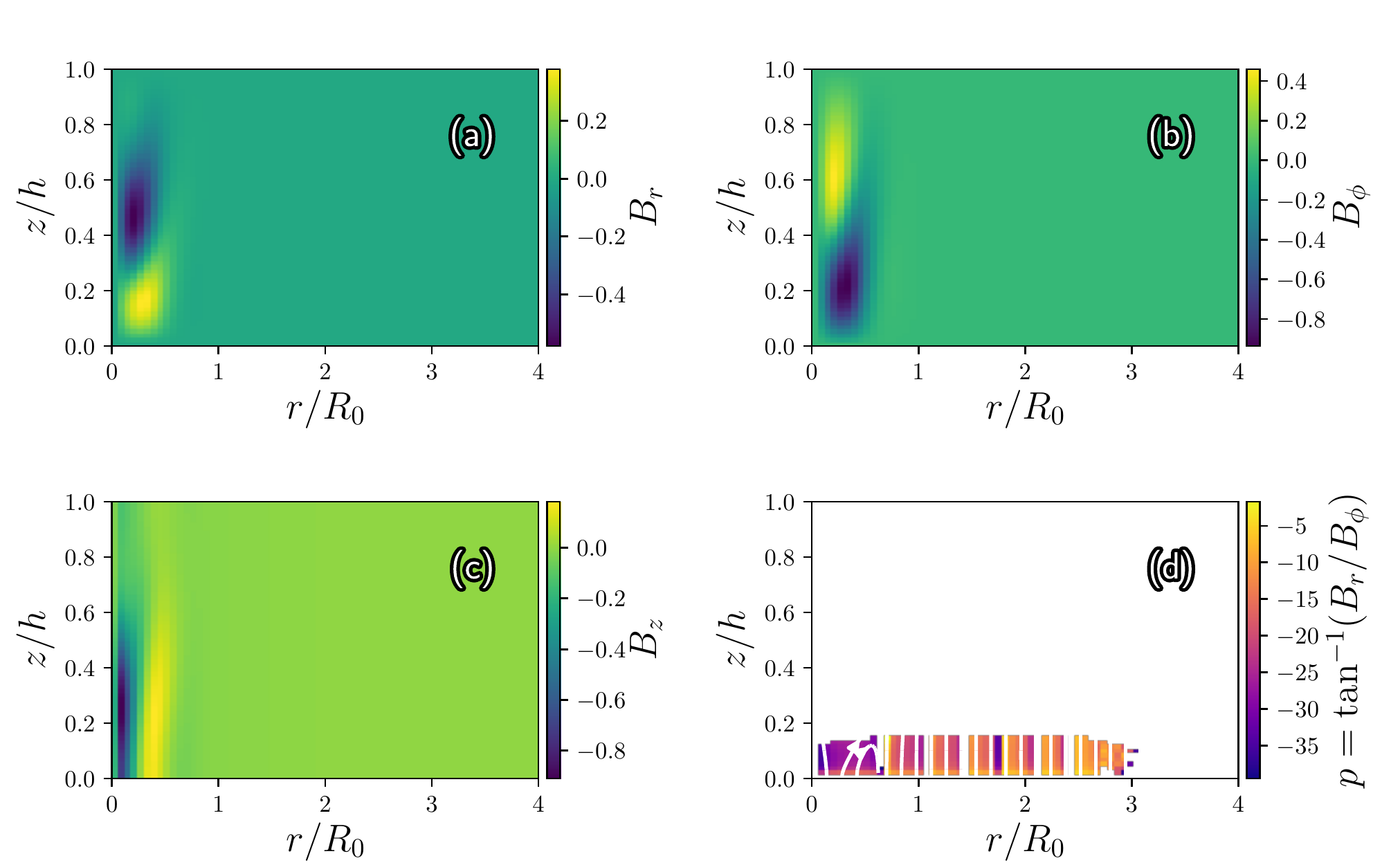}
    \caption{As in Fig.~\ref{fig:2dBcomponentsQuadrupolar} but for the dipolar parity. The critical circular rotation speed $V\crit/v_0=17$ is higher than for the marginal quadrupolar mode. Unlike the quadrupolar field of Fig.~\ref{fig:2dBcomponentsQuadrupolar}, the dipolar magnetic field varies periodically in time (see Fig.~\ref{fig:2dBcomponentsDipolarPhases}).
    }
    \label{fig:2dBcomponentsDipolar}
\end{figure}

\begin{figure}
	\includegraphics[width=\columnwidth]{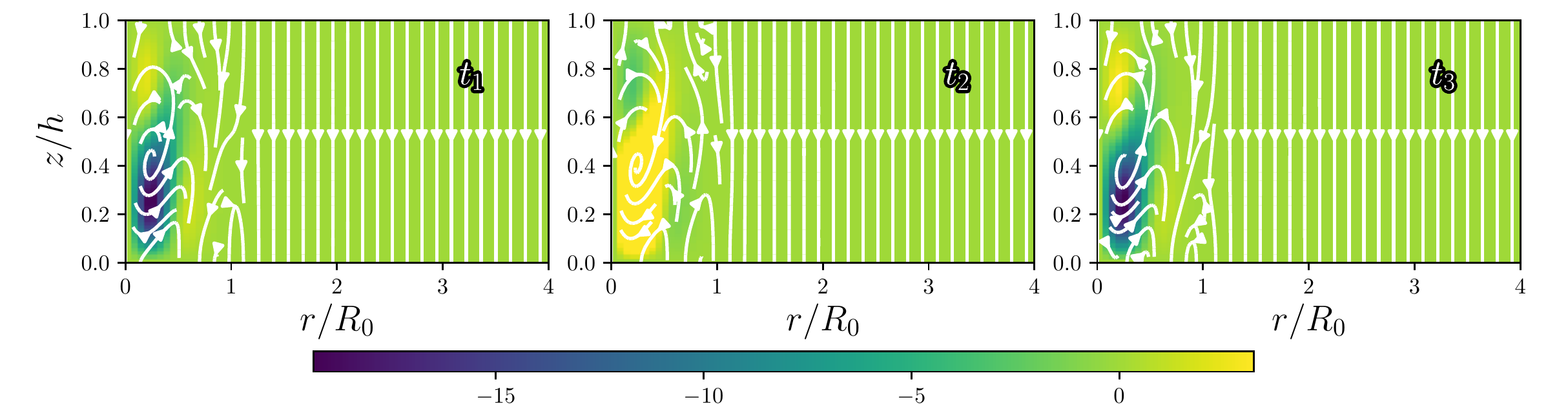}
    \caption{The form of the marginally stable dipolar mode of Fig.~\ref{fig:2dBcomponentsDipolar} in various phases of its oscillation: magnetic lines of the poloidal field are shown white and the toroidal component is colour-coded (arbitrary units) at three different times within the oscillation period ($t_1<t_2<t_3$).
    }
    \label{fig:2dBcomponentsDipolarPhases}
\end{figure}
\subsection{Magnetic field geometry}\label{MFG}
Large-scale magnetic fields of quadrupolar parity dominate in thin discs \citep{Ruzmaikin+1988p} as their critical dynamo number is lower than that for a dipolar magnetic field (or, equivalently, the magnetic field growth rate is higher for quadrupolar parity for a given dynamo number) and, as shown here, this difference is maintained in a thicker layer although it decreases as the disc becomes thicker ($\epsilon$ increases). In 
a sphere, dipolar and quadrupolar modes are generated as almost equal ease \citep{MREP77}. 

The structure of magnetic field obtained for $D=D\crit$ is shown in Figs~\ref{fig:2dBcomponentsQuadrupolar}--\ref{fig:2dBcomponentsDipolarPhases}. Typically of the eigenfunctions of the kinematic dynamo problem, magnetic field is localised near the maximum of the rotational shear \citep{RSSS90}. Magnetic pitch angles shown in Fig.~\ref{fig:2dBcomponentsQuadrupolar}d and \ref{fig:2dBcomponentsDipolar}d are similar to those observed in galaxies, ($-(10\text{--}30^\circ$), with negative values indicative of a spiral trailing with respect to the galactic rotation. Since the rotational shear in low-mass galaxies is lower than in massive spiral galaxies, the magnitude of the pitch angle is larger in the former, i.e., magnetic spirals are less tightly wound. Nonlinear dynamo effects, neglected here, lead to the saturation of the exponential growth of the magnetic field and to its spread along the disc. We do not discuss such nonlinear solutions here because they are more relevant in more realistic dynamo models that allow, in particular, for deviations of the discs of low-mass galaxies from axial symmetry.

\section{\label{CD}Discussion and conclusions}
Low-mass galaxies rotate slower than massive disc galaxies and hence the driving of the dynamo due to both rotational velocity shear and the $\alpha$-effect is weaker. The intensity of interstellar turbulence in low-mass galaxies is not lower  than in massive disc galaxies, so the turbulent magnetic diffusivity which destroys the mean magnetic field is similar in both massive and low-mass galaxies. We have shown that the ratio of the disc thickness to its radius plays a crucial role: mean-field dynamo can amplify and maintain magnetic fields in thicker discs at smaller rotational speeds than in thin discs because the turbulent diffusion time across the disc is longer in a thicker layer, and hence large-scale magnetic fields require lower rotational velocity shear to be maintained. Approximate solutions of kinematic $\alpha\omega$-dynamo equations applicable to thick discs obtained here clearly demonstrate an increase in the dynamo efficiency as the disc aspect ratio increases. The accuracy of the approximate solution is confirmed using numerical solutions of the dynamo equations. We conclude that mean-field dynamo can produce large-scale magnetic fields in low-mass galaxies.

We have only considered axially symmetric solutions of the kinematic dynamo equations. These are the main simplifications adopted in this work but they are appropriate when the origin of magnetic fields is considered rather than their form in specific galaxies. The discs of low-mass galaxies often exhibit strong deviations from axial symmetry, and this would affect the large-scale magnetic field. There are no reasons to expect that the mean-field dynamo action in a thick axisymmetric disc might support non-axisymmetric dynamo modes: they are difficult to maintain even in a thin disc, and the difficulty increases as the disc becomes thicker \citep{Ruzmaikin+1988p}. Therefore, the global magnetic structures observed in low-mass galaxies can be considered as a result of the distortion of a basic axially symmetric magnetic field by the deviations of the galaxy from axial asymmetry. Applications of dynamo theory to specific galaxies with allowance for deviations from axial symmetry and for the nonlinear saturation of the dynamo action will be discussed elsewhere.

\authorcontributions{conceptualization and methodology, A.S.\ and K.S.; 
software and analysis, P.B.;  writing -- review and editing, P.B., A.S.\ and K.S.}

\funding{A.S.\ acknowledges financial support of the STFC (ST/N000900/1, Project~2) and the Leverhulme Trust (RPG-2014-427).}

\acknowledgments{A.S.\ expresses gratitude to IUCAA for hospitality and financial support. K.S.\ is grateful to School of Mathematics, Statistics and Physics of Newcastle University for hospitality. P.B.\ thanks Arunima Benerjee, Eric Blackman, Luke Chamandy, Jayaram Chengalur, Dipanjan Mitra, Subhashis Roy for helpful comments.}

\conflictsofinterest{The authors declare no conflict of interest. The funders had no role in the design of the study; in the collection, analyses, or interpretation of data; in the writing of the manuscript, or in the decision to publish the results.} 


\reftitle{References}
\externalbibliography{yes}
\bibliography{dwarf}

\begin{thebibliography}{-------}
\providecommand{\natexlab}[1]{#1}

\bibitem[{Ruzmaikin} \em{et~al.}(1988){Ruzmaikin}, {Shukurov}, and
  {Sokoloff}]{Ruzmaikin+1988p}
{Ruzmaikin}, A.; {Shukurov}, A.; {Sokoloff}, D.
\newblock {\em {Magnetic Fields of Galaxies}}; Kluwer: Dordrecht,  1988.
\newblock
  doi:{\changeurlcolor{black}\href{https://doi.org/10.1007/978-94-009-2835-0}{\detokenize{10.1007/978-94-009-2835-0}}}.

\bibitem[{Beck} \em{et~al.}(1996){Beck}, {Brandenburg}, {Moss}, {Shukurov}, and
  {Sokoloff}]{Beck+1996}
{Beck}, R.; {Brandenburg}, A.; {Moss}, D.; {Shukurov}, A.; {Sokoloff}, D.
\newblock {Galactic magnetism: Recent developments and perspectives}.
\newblock {\em \araa} {\bf 1996}, {\em 34},~155--206.
\newblock
  doi:{\changeurlcolor{black}\href{https://doi.org/10.1146/annurev.astro.34.1.155}{\detokenize{10.1146/annurev.astro.34.1.155}}}.

\bibitem[{Beck}(2015)]{Beck15}
{Beck}, R.
\newblock {Magnetic fields in spiral galaxies}.
\newblock {\em \aapr} {\bf 2015}, {\em 24},~4.
\newblock
  doi:{\changeurlcolor{black}\href{https://doi.org/10.1007/s00159-015-0084-4}{\detokenize{10.1007/s00159-015-0084-4}}}.

\bibitem[{Brandenburg} and {Subramanian}(2005)]{Brandenburg+Subramanian2005}
{Brandenburg}, A.; {Subramanian}, K.
\newblock {Astrophysical magnetic fields and nonlinear dynamo theory}.
\newblock {\em \physrep} {\bf 2005}, {\em 417},~1--209,
  \href{http://xxx.lanl.gov/abs/astro-ph/0405052}{{\normalfont
  [astro-ph/0405052]}}.
\newblock
  doi:{\changeurlcolor{black}\href{https://doi.org/10.1016/j.physrep.2005.06.005}{\detokenize{10.1016/j.physrep.2005.06.005}}}.

\bibitem[{Shukurov}(2007)]{Sh07}
{Shukurov}, A., {Galactic dynamos}.
\newblock In {\em Mathematical Aspects of Natural Dynamos}; {Dormy}, E.;
  {Soward}, A.M., Eds.; Chapman \& Hall/CRC,  2007; pp. 313--359.

\bibitem[{Chamandy} \em{et~al.}(2014){Chamandy}, {Shukurov}, {Subramanian}, and
  {Stoker}]{CSSS14}
{Chamandy}, L.; {Shukurov}, A.; {Subramanian}, K.; {Stoker}, K.
\newblock {Non-linear galactic dynamos: a toolbox}.
\newblock {\em \mnras} {\bf 2014}, {\em 443},~1867--1880.
\newblock
  doi:{\changeurlcolor{black}\href{https://doi.org/10.1093/mnras/stu1274}{\detokenize{10.1093/mnras/stu1274}}}.

\bibitem[{Shukurov} and {Subramanian}(2019)]{SS19}
{Shukurov}, A.; {Subramanian}, K.
\newblock {\em {Astrophysical Magnetic Fields: From Galaxies to the Early
  Universe}}; CUP: Cambridge,  2019.

\bibitem[{Beck} \em{et~al.}(2005){Beck}, {Fletcher}, {Shukurov}, {Snodin},
  {Sokoloff}, {Ehle}, {Moss}, and {Shoutenkov}]{BFSSSEMS05}
{Beck}, R.; {Fletcher}, A.; {Shukurov}, A.; {Snodin}, A.; {Sokoloff}, D.D.;
  {Ehle}, M.; {Moss}, D.; {Shoutenkov}, V.
\newblock {Magnetic fields in barred galaxies. IV. NGC~1097 and NGC~1365}.
\newblock {\em \aap} {\bf 2005}, {\em 444},~739--765,
  \href{http://xxx.lanl.gov/abs/astro-ph/0508485}{{\normalfont
  [astro-ph/0508485]}}.
\newblock
  doi:{\changeurlcolor{black}\href{https://doi.org/10.1051/0004-6361:20053556}{\detokenize{10.1051/0004-6361:20053556}}}.

\bibitem[{Van Eck} \em{et~al.}(2015){Van Eck}, {Brown}, {Shukurov}, and
  {Fletcher}]{VEBSF15}
{Van Eck}, C.L.; {Brown}, J.C.; {Shukurov}, A.; {Fletcher}, A.
\newblock {Magnetic fields in a sample of nearby spiral galaxies}.
\newblock {\em \apj} {\bf 2015}, {\em 799},~35.
\newblock
  doi:{\changeurlcolor{black}\href{https://doi.org/10.1088/0004-637X/799/1/35}{\detokenize{10.1088/0004-637X/799/1/35}}}.

\bibitem[{Chamandy} \em{et~al.}(2016){Chamandy}, {Shukurov}, and
  {Taylor}]{CST16}
{Chamandy}, L.; {Shukurov}, A.; {Taylor}, A.R.
\newblock {Statistical tests of galactic dynamo theory}.
\newblock {\em \apj} {\bf 2016}, {\em 833},~43.
\newblock
  doi:{\changeurlcolor{black}\href{https://doi.org/10.3847/1538-4357/833/1/43}{\detokenize{10.3847/1538-4357/833/1/43}}}.

\bibitem[{Rodrigues} \em{et~al.}(2015){Rodrigues}, {Shukurov}, {Fletcher}, and
  {Baugh}]{RSFB15}
{Rodrigues}, L.F.S.; {Shukurov}, A.; {Fletcher}, A.; {Baugh}, C.M.
\newblock {Galactic magnetic fields and hierarchical galaxy formation}.
\newblock {\em \mnras} {\bf 2015}, {\em 450},~3472--3489.
\newblock
  doi:{\changeurlcolor{black}\href{https://doi.org/10.1093/mnras/stv816}{\detokenize{10.1093/mnras/stv816}}}.

\bibitem[{Rodrigues} \em{et~al.}(2019){Rodrigues}, {Chamandy}, {Shukurov},
  {Baugh}, and {Taylor}]{RCSBT19}
{Rodrigues}, L.F.S.; {Chamandy}, L.; {Shukurov}, A.; {Baugh}, C.M.; {Taylor},
  A.R.
\newblock {Evolution of galactic magnetic fields}.
\newblock {\em \mnras} {\bf 2019}, {\em 483},~2424--2440.
\newblock
  doi:{\changeurlcolor{black}\href{https://doi.org/10.1093/mnras/sty3270}{\detokenize{10.1093/mnras/sty3270}}}.

\bibitem[{Chy{\.z}y} \em{et~al.}(2000){Chy{\.z}y}, {Beck}, {Kohle}, {Klein},
  and {Urbanik}]{Chyzy+2000}
{Chy{\.z}y}, K.T.; {Beck}, R.; {Kohle}, S.; {Klein}, U.; {Urbanik}, M.
\newblock {Regular magnetic fields in the dwarf irregular galaxy NGC~4449}.
\newblock {\em \aap} {\bf 2000}, {\em 355},~128--137,
  \href{http://xxx.lanl.gov/abs/astro-ph/0001205}{{\normalfont
  [astro-ph/0001205]}}.

\bibitem[{Haynes} \em{et~al.}(1991){Haynes}, {Klein}, {Wayte}, {Wielebinski},
  {Murray}, {Bajaja}, {Meinert}, {Buczilowski}, {Harnett}, {Hunt}, {Wark}, and
  {Sciacca}]{H+91}
{Haynes}, R.F.; {Klein}, U.; {Wayte}, S.R.; {Wielebinski}, R.; {Murray}, J.D.;
  {Bajaja}, E.; {Meinert}, D.; {Buczilowski}, U.R.; {Harnett}, J.I.; {Hunt},
  A.J.; {Wark}, R.; {Sciacca}, L.
\newblock {A radio continuum study of the Magellanic Clouds. I. Complete
  multi-frequency maps}.
\newblock {\em \aap} {\bf 1991}, {\em 252},~475--486.

\bibitem[{Mao} \em{et~al.}(2008){Mao}, {Gaensler}, {Stanimirovi{\'c}},
  {Haverkorn}, {McClure-Griffiths}, {Staveley- Smith}, and {Dickey}]{Mao+08}
{Mao}, S.A.; {Gaensler}, B.M.; {Stanimirovi{\'c}}, S.; {Haverkorn}, M.;
  {McClure-Griffiths}, N.M.; {Staveley- Smith}, L.; {Dickey}, J.M.
\newblock {A radio and optical polarization study of the magnetic field in the
  Small Magellanic Cloud}.
\newblock {\em \apj} {\bf 2008}, {\em 688},~1029--1049.
\newblock
  doi:{\changeurlcolor{black}\href{https://doi.org/10.1086/590546}{\detokenize{10.1086/590546}}}.

\bibitem[{Mao} \em{et~al.}(2012){Mao}, {McClure-Griffiths}, {Gaensler},
  {Haverkorn}, {Beck}, {McConnell}, {Wolleben}, {Stanimirovi{\'c}}, {Dickey},
  and {Staveley-Smith}]{Mao+12}
{Mao}, S.A.; {McClure-Griffiths}, N.M.; {Gaensler}, B.M.; {Haverkorn}, M.;
  {Beck}, R.; {McConnell}, D.; {Wolleben}, M.; {Stanimirovi{\'c}}, S.;
  {Dickey}, J.M.; {Staveley-Smith}, L.
\newblock {Magnetic field structure of the Large Magellanic Cloud from Faraday
  rotation measures of diffuse polarized emission}.
\newblock {\em \apj} {\bf 2012}, {\em 759},~25.
\newblock
  doi:{\changeurlcolor{black}\href{https://doi.org/10.1088/0004-637X/759/1/25}{\detokenize{10.1088/0004-637X/759/1/25}}}.

\bibitem[{Kepley} \em{et~al.}(2010){Kepley}, {M{\"u}hle}, {Everett}, {Zweibel},
  {Wilcots}, and {Klein}]{Kepley+10}
{Kepley}, A.A.; {M{\"u}hle}, S.; {Everett}, J.; {Zweibel}, E.G.; {Wilcots},
  E.M.; {Klein}, U.
\newblock {The role of the magnetic field in the interstellar medium of the
  post-starburst dwarf irregular galaxy NGC~1569}.
\newblock {\em \apj} {\bf 2010}, {\em 712},~536--557.
\newblock
  doi:{\changeurlcolor{black}\href{https://doi.org/10.1088/0004-637X/712/1/536}{\detokenize{10.1088/0004-637X/712/1/536}}}.

\bibitem[{Chy{\.z}y} \em{et~al.}(2011){Chy{\.z}y}, {We{\.z}gowiec}, {Beck}, and
  {Bomans}]{Chyzy+2011}
{Chy{\.z}y}, K.T.; {We{\.z}gowiec}, M.; {Beck}, R.; {Bomans}, D.J.
\newblock {Magnetic fields in Local Group dwarf irregulars}.
\newblock {\em \aap} {\bf 2011}, {\em 529},~A94.
\newblock
  doi:{\changeurlcolor{black}\href{https://doi.org/10.1051/0004-6361/201015393}{\detokenize{10.1051/0004-6361/201015393}}}.

\bibitem[{Hindson} \em{et~al.}(2018){Hindson}, {Kitchener}, {Brinks}, {Heesen},
  {Westcott}, {Hunter}, {Zhang}, {Rupen}, and {Rau}]{HKBHWHZRR18}
{Hindson}, L.; {Kitchener}, G.; {Brinks}, E.; {Heesen}, V.; {Westcott}, J.;
  {Hunter}, D.; {Zhang}, H.X.; {Rupen}, M.; {Rau}, U.
\newblock {A radio continuum study of dwarf galaxies: 6~cm imaging of LITTLE
  THINGS}.
\newblock {\em \apjs} {\bf 2018}, {\em 234},~29.
\newblock
  doi:{\changeurlcolor{black}\href{https://doi.org/10.3847/1538-4365/aaa42c}{\detokenize{10.3847/1538-4365/aaa42c}}}.

\bibitem[{Beck} and {Wielebinski}(2013)]{BW13}
{Beck}, R.; {Wielebinski}, R., {Magnetic fields in galaxies}.
\newblock In {\em Planets, Stars and Stellar Systems}; {Oswalt}, T.D.;
  {Gilmore}, G., Eds.; Springer: Dordrecht,  2013; Vol.~5, p. 641.
\newblock
  doi:{\changeurlcolor{black}\href{https://doi.org/10.1007/978-94-007-5612-0_13}{\detokenize{10.1007/978-94-007-5612-0_13}}}.

\bibitem[{Walter} and {Brinks}(1999)]{WaBr99}
{Walter}, F.; {Brinks}, E.
\newblock {Holes and shells in the interstellar medium of the nearby dwarf
  galaxy IC~2574}.
\newblock {\em \aj} {\bf 1999}, {\em 118},~273--301,
  \href{http://xxx.lanl.gov/abs/astro-ph/9904002}{{\normalfont
  [astro-ph/9904002]}}.
\newblock
  doi:{\changeurlcolor{black}\href{https://doi.org/10.1086/300906}{\detokenize{10.1086/300906}}}.

\bibitem[{Wilcots} and {Miller}(1998)]{Wilcots+Miller1998}
{Wilcots}, E.M.; {Miller}, B.W.
\newblock {The kinematics and distribution of H\,\textsc{i} in IC~10}.
\newblock {\em \aj} {\bf 1998}, {\em 116},~2363--2394.
\newblock
  doi:{\changeurlcolor{black}\href{https://doi.org/10.1086/300595}{\detokenize{10.1086/300595}}}.

\bibitem[{Oh} \em{et~al.}(2015){Oh}, {Hunter}, {Brinks}, {Elmegreen},
  {Schruba}, {Walter}, {Rupen}, {Young}, {Simpson}, {Johnson}, {Herrmann},
  {Ficut-Vicas}, {Cigan}, {Heesen}, {Ashley}, and {Zhang}]{Oh+15}
{Oh}, S.H.; {Hunter}, D.A.; {Brinks}, E.; {Elmegreen}, B.G.; {Schruba}, A.;
  {Walter}, F.; {Rupen}, M.P.; {Young}, L.M.; {Simpson}, C.E.; {Johnson}, M.C.;
  {Herrmann}, K.A.; {Ficut-Vicas}, D.; {Cigan}, P.; {Heesen}, V.; {Ashley}, T.;
  {Zhang}, H.X.
\newblock {High-resolution mass models of dwarf galaxies from LITTLE THINGS}.
\newblock {\em \aj} {\bf 2015}, {\em 149},~180.
\newblock
  doi:{\changeurlcolor{black}\href{https://doi.org/10.1088/0004-6256/149/6/180}{\detokenize{10.1088/0004-6256/149/6/180}}}.

\bibitem[{Iorio} \em{et~al.}(2017){Iorio}, {Fraternali}, {Nipoti}, {Di
  Teodoro}, {Read}, and {Battaglia}]{IFNDTRB17}
{Iorio}, G.; {Fraternali}, F.; {Nipoti}, C.; {Di Teodoro}, E.; {Read}, J.I.;
  {Battaglia}, G.
\newblock {LITTLE THINGS in 3D: robust determination of the circular velocity
  of dwarf irregular galaxies}.
\newblock {\em \mnras} {\bf 2017}, {\em 466},~4159--4192.
\newblock
  doi:{\changeurlcolor{black}\href{https://doi.org/10.1093/mnras/stw3285}{\detokenize{10.1093/mnras/stw3285}}}.

\bibitem[{Martimbeau} \em{et~al.}(1994){Martimbeau}, {Carignan}, and
  {Roy}]{MCR94}
{Martimbeau}, N.; {Carignan}, C.; {Roy}, J.R.
\newblock {Dark matter distribution and the H\,\textsc{i}--H$\alpha$ connection
  in IC~2574}.
\newblock {\em \aj} {\bf 1994}, {\em 107},~543--554.
\newblock
  doi:{\changeurlcolor{black}\href{https://doi.org/10.1086/116875}{\detokenize{10.1086/116875}}}.

\bibitem[{Staveley-Smith} \em{et~al.}(1992){Staveley-Smith}, {Davies}, and
  {Kinman}]{S-SDK92}
{Staveley-Smith}, L.; {Davies}, R.D.; {Kinman}, T.D.
\newblock {H\,\textsc{i} and optical observations of dwarf galaxies}.
\newblock {\em \mnras} {\bf 1992}, {\em 258},~334--346.
\newblock
  doi:{\changeurlcolor{black}\href{https://doi.org/10.1093/mnras/258.2.334}{\detokenize{10.1093/mnras/258.2.334}}}.

\bibitem[{Roychowdhury} \em{et~al.}(2013){Roychowdhury}, {Chengalur},
  {Karachentsev}, and {Kaisina}]{Roychowdhury+2013}
{Roychowdhury}, S.; {Chengalur}, J.N.; {Karachentsev}, I.D.; {Kaisina}, E.I.
\newblock {The intrinsic shapes of dwarf irregular galaxies}.
\newblock {\em \mnras} {\bf 2013}, {\em 436},~L104--L108.
\newblock
  doi:{\changeurlcolor{black}\href{https://doi.org/10.1093/mnrasl/slt123}{\detokenize{10.1093/mnrasl/slt123}}}.

\bibitem[{Banerjee} \em{et~al.}(2011){Banerjee}, {Jog}, {Brinks}, and
  {Bagetakos}]{BaJoBrBa11}
{Banerjee}, A.; {Jog}, C.J.; {Brinks}, E.; {Bagetakos}, I.
\newblock {Theoretical determination of H\,\textsc{i} vertical scale heights in
  the dwarf galaxies DDO~154, Ho~II, IC~2574 and NGC~2366}.
\newblock {\em \mnras} {\bf 2011}, {\em 415},~687--694.
\newblock
  doi:{\changeurlcolor{black}\href{https://doi.org/10.1111/j.1365-2966.2011.18745.x}{\detokenize{10.1111/j.1365-2966.2011.18745.x}}}.

\bibitem[{Zeldovich} \em{et~al.}(1990){Zeldovich}, {Ruzmaikin}, and
  {Sokoloff}]{ZRS90}
{Zeldovich}, Y.B.; {Ruzmaikin}, A.A.; {Sokoloff}, D.D.
\newblock {\em {The Almighty Chance}}; World Scientific: Singapore,  1990.
\newblock
  doi:{\changeurlcolor{black}\href{https://doi.org/10.1142/0862}{\detokenize{10.1142/0862}}}.

\bibitem[{Ivers}(2017)]{Ivers17}
{Ivers}, D.J.
\newblock {Kinematic dynamos in spheroidal geometries}.
\newblock {\em Proc.\ Roy.\ Soc.\ London, Ser.\ A} {\bf 2017}, {\em
  473},~20170432.
\newblock
  doi:{\changeurlcolor{black}\href{https://doi.org/10.1098/rspa.2017.0432}{\detokenize{10.1098/rspa.2017.0432}}}.

\bibitem[{Siejkowski} \em{et~al.}(2018){Siejkowski}, {Soida}, and
  {Chy{\.z}y}]{SiSoCh18}
{Siejkowski}, H.; {Soida}, M.; {Chy{\.z}y}, K.T.
\newblock {Magnetic field evolution in dwarf and Magellanic-type galaxies}.
\newblock {\em \aap} {\bf 2018}, {\em 611},~A7.
\newblock
  doi:{\changeurlcolor{black}\href{https://doi.org/10.1051/0004-6361/201730566}{\detokenize{10.1051/0004-6361/201730566}}}.

\bibitem[{Siejkowski} \em{et~al.}(2010){Siejkowski}, {Soida},
  {Otmianowska-Mazur}, {Hanasz}, and {Bomans}]{Siejkowski+2010}
{Siejkowski}, H.; {Soida}, M.; {Otmianowska-Mazur}, K.; {Hanasz}, M.; {Bomans},
  D.J.
\newblock {Cosmic-ray driven dynamo in the interstellar medium of irregular
  galaxies}.
\newblock {\em \aap} {\bf 2010}, {\em 510},~A97.
\newblock
  doi:{\changeurlcolor{black}\href{https://doi.org/10.1051/0004-6361/200912729}{\detokenize{10.1051/0004-6361/200912729}}}.

\bibitem[{Siejkowski} \em{et~al.}(2014){Siejkowski}, {Otmianowska-Mazur},
  {Soida}, {Bomans}, and {Hanasz}]{SO-MSBH14}
{Siejkowski}, H.; {Otmianowska-Mazur}, K.; {Soida}, M.; {Bomans}, D.J.;
  {Hanasz}, M.
\newblock {3D global simulations of a cosmic-ray-driven dynamo in dwarf
  galaxies}.
\newblock {\em \aap} {\bf 2014}, {\em 562},~A136.
\newblock
  doi:{\changeurlcolor{black}\href{https://doi.org/10.1051/0004-6361/201220367}{\detokenize{10.1051/0004-6361/201220367}}}.

\bibitem[{Moffatt}(1978)]{Moffatt78}
{Moffatt}, H.K.
\newblock {\em {Magnetic Field Generation in Electrically Conducting Fluids}};
  CUP: Cambridge,  1978.

\bibitem[{Parker}(1979)]{Parker79}
{Parker}, E.N.
\newblock {\em {Cosmical Magnetic Fields. Their Origin and Their Activity}};
  Clarendon Press: Oxford,  1979.

\bibitem[{Krause} and {R\"adler}(1980)]{KR80}
{Krause}, F.; {R\"adler}, K.H.
\newblock {\em {Mean-field Magnetohydrodynamics and Dynamo Theory}}; Pergamon
  Press: Oxford,  1980.

\bibitem[{Zeldovich} \em{et~al.}(1983){Zeldovich}, {Ruzmaikin}, and
  {Sokoloff}]{ZRS83}
{Zeldovich}, {\relax Ya}.B.; {Ruzmaikin}, A.A.; {Sokoloff}, D.D.
\newblock {\em {Magnetic Fields in Astrophysics}}; Gordon and Breach: New York,
   1983.

\bibitem[{Shukurov} \em{et~al.}(2019){Shukurov}, {Rodrigues}, {Bushby},
  {Hollins}, and {Rachen}]{SRBHR18}
{Shukurov}, A.; {Rodrigues}, L.F.S.; {Bushby}, P.J.; {Hollins}, J.; {Rachen},
  J.P.
\newblock {A physical approach to modelling large-scale galactic magnetic
  fields}.
\newblock {\em \aap} {\bf 2019}, {\em 623},~A113,
  \href{http://xxx.lanl.gov/abs/1809.03595}{{\normalfont
  [arXiv:astro-ph.GA/1809.03595]}}.
\newblock
  doi:{\changeurlcolor{black}\href{https://doi.org/10.1051/0004-6361/201834642}{\detokenize{10.1051/0004-6361/201834642}}}.

\bibitem[{Subramanian} and {Mestel}(1993)]{SM93}
{Subramanian}, K.; {Mestel}, L.
\newblock {Galactic dynamos and density wave theory - II. an alternative
  treatment for strong non-axisymmetry.}
\newblock {\em \mnras} {\bf 1993}, {\em 265},~649--654.
\newblock
  doi:{\changeurlcolor{black}\href{https://doi.org/10.1093/mnras/265.3.649}{\detokenize{10.1093/mnras/265.3.649}}}.

\bibitem[{Phillips}(2001)]{P01}
{Phillips}, A.
\newblock {A comparison of the asymptotic and no-{$z$} approximations for
  galactic dynamos}.
\newblock {\em Geophys.\ Astrophys.\ Fluid Dyn.} {\bf 2001}, {\em
  94},~135--150.
\newblock
  doi:{\changeurlcolor{black}\href{https://doi.org/10.1080/03091920108204133}{\detokenize{10.1080/03091920108204133}}}.

\bibitem[{Ji} \em{et~al.}(2014){Ji}, {Cole}, {Bushby}, and {Shukurov}]{JCBS14}
{Ji}, Y.; {Cole}, L.; {Bushby}, P.; {Shukurov}, A.
\newblock {Asymptotic solutions for mean-field slab dynamos}.
\newblock {\em Geophys.\ Astrophys.\ Fluid Dyn.} {\bf 2014}, {\em
  108},~568--583.
\newblock
  doi:{\changeurlcolor{black}\href{https://doi.org/10.1080/03091929.2014.898757}{\detokenize{10.1080/03091929.2014.898757}}}.

\bibitem[{Shukurov} and {Sokoloff}(2008)]{ShSo08}
{Shukurov}, A.; {Sokoloff}, D., {Astrophysical dynamos}.
\newblock In {\em Dynamos: Lecture Notes of the Les Houches Summer School
  2007}; Cardin, P.; Cugliandolo, L., Eds.; Elsevier Science,  2008; pp.
  251--299.

\bibitem[{R{\"a}dler} and {Wiedemann}(1989)]{RaWi89}
{R{\"a}dler}, K.H.; {Wiedemann}, E.
\newblock {Numerical experiments with a simple nonlinear mean-field dynamo
  model}.
\newblock {\em Geophys.\ Astrophys.\ Fluid Dyn.} {\bf 1989}, {\em 49},~71--79.
\newblock
  doi:{\changeurlcolor{black}\href{https://doi.org/10.1080/03091928908243464}{\detokenize{10.1080/03091928908243464}}}.

\bibitem[{Proctor}(1977)]{MREP77}
{Proctor}, M.R.E.
\newblock {On the eigenvalues of kinematic {\ensuremath{\alpha}}-effect
  dynamos}.
\newblock {\em Astron.\ Nachr.} {\bf 1977}, {\em 298},~19.
\newblock
  doi:{\changeurlcolor{black}\href{https://doi.org/10.1002/asna.19772980104}{\detokenize{10.1002/asna.19772980104}}}.

\bibitem[{Ruzmaikin} \em{et~al.}(1990){Ruzmaikin}, {Shukurov}, {Sokoloff}, and
  {Starchenko}]{RSSS90}
{Ruzmaikin}, A.; {Shukurov}, A.; {Sokoloff}, D.; {Starchenko}, S.
\newblock {Maximally-efficient-generation approach in the dynamo theory}.
\newblock {\em Geophys.\ Astrophys.\ Fluid Dyn.} {\bf 1990}, {\em
  52},~125--139.
\newblock
  doi:{\changeurlcolor{black}\href{https://doi.org/10.1080/03091929008219843}{\detokenize{10.1080/03091929008219843}}}.

\end{thebibliography}
\label{lastpage}
\end{document}